\documentstyle[11pt,newpasp,twoside,epsf]{article}
\def\edcomment#1{\iffalse\marginpar{\raggedright\sl#1\/}\else\relax\fi}
\reversemarginpar

\begin{document}
\title{Issues in Blazar Research}
\author{Paolo Padovani}
\affil{Space Telescope Science Institute, 3700 San Martin Drive, Baltimore,
MD, 21218, USA}
\affil{Affiliated to the Astrophysics Division, Space Science Department, European
Space Agency}
\affil{On leave from Dipartimento di Fisica, II Universit\`a di Roma ``Tor Vergata,''
Via della Ricerca Scientifica~1, I-00133 Roma, Italy}
\author{C.\ Megan Urry}
\affil{Space Telescope Science Institute, 3700 San Martin Drive, Baltimore,
MD, 21218, USA} 

\begin{abstract}
As an overview of blazar research, we briefly discuss some of the themes of this conference, including the various types of blazars, blazar samples and selection effects, unified schemes, evolution, and jet physics.
\end{abstract}

\section{Introduction}

The initial motivation for this conference was to try to solve the seemingly simple problem ``which blazars are more numerous,'' HBL-like or LBL-like. We thought that by convening in the same room advocates from the various factions we could perhaps agree on an answer. As this question had broad implications for our understanding of blazars, we expanded the scope of the conference to include also jet physics. The timing was good. After years during which our knowledge of blazars was based mostly on relatively shallow and small samples, new, deeper samples are now being assembled out of the large surveys that are available at various wavelengths. 

More than fifty astronomers responded to our call. Many of them were the ``usual suspects'' but also a good number were young scientists, relatively new to the field, a sign that blazars continue to arouse interest.

The conference program addressed some of the outstanding key issues in blazar research. It is impossible to do justice to the whole conference in a short paper. Instead we highlight some of the most fundamental themes, referring to the single papers for a more in-depth discussion. We hope in this way to give a glimpse of the ``big picture'' underlying the various efforts.

One demographic issue we clearly solved, at least, was that of nationality. We had participants from all over the world and we were curious to see ``which blazar pundits are more numerous.'' The answer should have been obvious to anybody overhearing some of the conversations during the coffee breaks but it comes out explicitly in Fig.~1: blazars are Italian!

\begin{figure}
\centerline{\epsfxsize=9.0cm 
\epsfbox{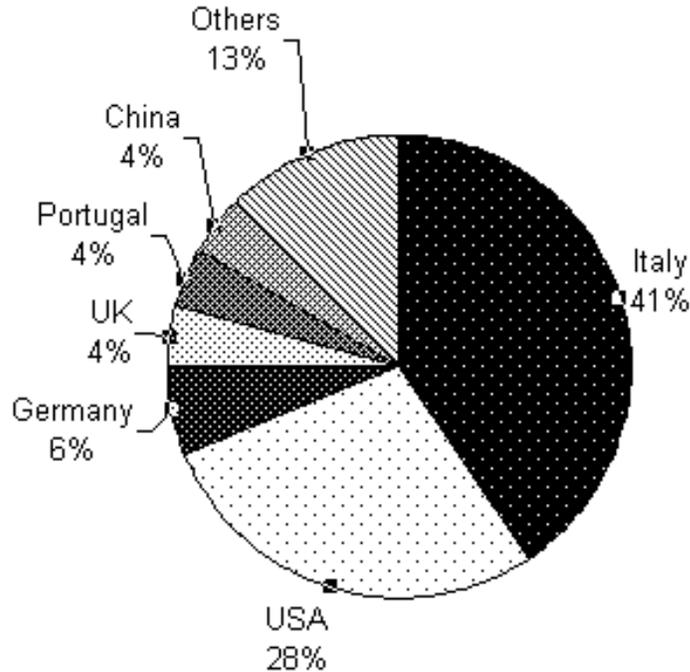}} 
\caption{The distribution of nationalities amongst the attendants to the blazar conference.}
\end{figure}

\section{Types of Blazars}

There is general agreement that blazars are radio-loud active galactic nuclei with relativistic jets pointed at us, which means their numbers and jet physics are relevant to all radio-loud active galactic nuclei (AGN). Blazars include both BL~Lac objects, which have very weak line emission, and more luminous flat-spectrum radio quasars (FSRQ), which have normal, strong emission lines. Notwithstanding that nomenclature is the bane of this field, the conference necessarily began with definitions of common terms for types of blazars, RBL and XBL, LBL and HBL, red and blue. Because BL~Lacs found in radio and X-ray surveys had strikingly different properties, they were further designated radio-selected BL~Lac (RBL) or 
X-ray-selected (XBL) BL~Lac. It was eventually realized that most RBL were radio bright and X-ray weak because their synchrotron component peaked at relatively low frequency (infrared-optical), while most XBL were radio weak and X-ray bright because
their peaks fell at ultraviolet-X-ray frequencies. Hence the more precise terms: 
low-frequency peaked (LBL) and high-frequency peaked (HBL) BL~Lacs, depending on whether the value of the two-point radio/X-ray spectral index $\alpha_{\rm rx}$ is greater than or less than $\sim0.8$, respectively. These definitions extend naturally to all blazars, both BL~Lacs and quasars. Finally, to avoid the over-use of these undescriptive acronyms, we often employ the more colorful terms ``red'' and ``blue'' blazars for LBL and HBL, respectively.

Several participants emphasized that blazars show a clear continuity of properties with radio galaxies and steep-spectrum radio quasars (Scarpa, Mara\-schi, Anton, Wolter, March\~a, Ma). Radio powers and emission line strengths rise smoothly from 
BL~Lacs to quasars---the very definition of a BL~Lac as having equivalent width emission lines weaker than 5~\AA\ is completely arbitrary (Landt). In a similar way, the radio powers and host galaxy magnitudes of the ``parent population'' of blazars, radio galaxies of Fanaroff-Riley (FR) types~I and II, overlap. The host galaxies of blazars are uniformly luminous giant ellipticals, regardless of intrinsic nuclear power (Scarpa, Pursimo). They also have similar properties (morphologies, luminosities, and sizes) to the host galaxies of FR~I and IIs or to luminous ellipticals without central radio sources.

The spectral energy distributions (SED) of blazars were interpreted as a combination of synchrotron emission and higher-energy radiation, possibly from Compton scattering (Maraschi). The observed SEDs follow an interesting trend with luminosity, such that high luminosity blazars have synchrotron and Compton peaks at lower frequencies. This can be explained as a cooling sequence, although there are concerns that it is induced at least in part by strong selection effects in the limited samples studied. 

\section{Blazar Samples and Selection Effects}

The importance of selection biases was a recurring theme of this conference. The unification paradigm first arose a decade ago to explain the very strongest selection biases, that blazar radiation is strongly beamed in our direction (but maybe not as strongly as we think? Abraham), causing AGN appearance to change dramatically with orientation.  Now experts in the field are focusing on more subtle effects stemming from the SED shapes (implying strong K-corrections) and from variability (especially important at gamma-ray energies where detectors/telescopes are not yet very sensitive).

Most of our current knowledge of blazars is based on a handful of relatively small and high-flux-limit samples. As a result, we have only probed the intrinsically most luminous sources. The new, deeper surveys currently underway are changing that. Participants at this meeting described several major new efforts, including the BLEIS, DXRBS, FIRST, Parkes 0.25~Jy, REX, RGB, and Sedentary surveys (Cagnoni, Padovani, Caccianiga, Giommi). Many of these new surveys employ multiple flux limits, usually radio and X-ray, and in some cases also optical. Because of the steepness of the source counts (more faint sources), such surveys find objects mostly at the survey vertex in the flux-flux(-flux) plane (or volume). Extension of the results of any survey to the whole blazar population is not always possible and in any case always requires understanding how such flux limits bias the derived sample (Padovani).

The selection biases can be very strong. The distinction between HBL and LBL, which previously looked like different types of blazar, has now been shown to be far more arbitrary because many intermediate blazars have been found (e.g.\ in the RGB, DXRBS, and REX surveys). There must therefore be a broad distribution of peak synchrotron frequencies between the far-infrared and the X-ray bands. These new surveys have also found a previously unknown class of blue FSRQ, although there are clearly none as blue as the most extreme BL~Lacs (Perlman, Padovani, Wolter, Costamante). We still do not know how these blue FSRQ relate to their red relatives, although there are some ideas (Georganopoulos, Perlman). Simulations presented at the conference showed 
clearly that the content of new blazar samples, even the deeper ones, is affected more strongly by the sample flux limit(s) than by intrinsic properties (Fossati, Giommi). {\it In the simplest terms, the surveys find what they can rather than what exists.}

\section{Unification of Blazars and Radio Galaxies}

The original unification ideas, equating BL Lac objects with FR~I galaxies and quasars with FR~IIs, seem to hold up well to newer, more extensive data. The radio counts of BL~Lac objects from the $\sim20$-times deeper DXRBS survey match well the predictions from beaming models matched to the 34-object 1~Jy sample (i.e.\ starting from FR~I luminosity function). Similarly, the luminosity function of flat-spectrum radio quasars has now been extended a factor $\ga 20$ lower in luminosity by the DXRBS, and it too matches well predictions based on beaming FR~IIs into 
high-luminosity flat-spectrum radio quasars based on the 2~Jy sample (Padovani). Given the overlapping properties of FR~Is and FR~IIs, and of BL~Lacs and quasars, it seems likely there is one over-arching ``grand unification'' scheme relating all radio galaxies to all blazars, although we still do not know what this scheme is. Bulk Lorentz factors $\sim 10$ still explain, at zeroth order, the number statistics, the SED (Chiaberge), superluminal motion (Marscher), and the correlation between
core-dominance parameter and optical polarization (Yuan), although there might be some inconsistencies among different methods (Chiaberge).

A few intriguing details remain unresolved, however. Recent work with HST implies that nuclear obscuration (e.g.\ a torus) may be very rare in FR~I radio galaxies, implying that few have hidden broad-emission-line regions (Chiaberge). This would make them physically different from higher-power radio sources, where the presence of an obscuring torus has been inferred from spectropolarimetric observations (broad lines are seen in the polarized light from FR~II radio galaxies, presumably scattered from a hidden broad-line region). It would further imply that broad lines should be extremely rare in BL~Lac objects, yet they have been observed, including in BL~Lac itself. On the other hand, the spectral energy distribution of (at least a couple) 
FR~Is is consistent with that of BL~Lacs of the LBL type (Trussoni). 

We now fold evolution into the unification picture. The evolutionary properties of 
BL~Lac objects have been puzzling for a number of years. It appeared that the evolution of different samples was quite different, the radio-selected 1~Jy sample evolving in the same sense as quasars (more/brighter at high redshift) and the 
X-ray-selected EMSS sample evolving in the opposite sense. Participants at the conference brought some welcome clarity to this mess. In general, BL~Lac objects evolve little or not at all, regardless of the survey in which they were found (Padovani, Caccianiga, Stocke). There is, however, an intriguing dependence of evolution (as measured by $\langle V_e / V_a \rangle$, the ratio of ``enclosed'' to ``available'' volume) on spectral type, with the extreme HBL (very low $\alpha_{\rm rx}$ or equivalently very high $f_{\rm x}/f_{\rm r}$) still showing negative evolution (fewer/dimmer at high redshift; Giommi, Stocke). This could possibly be
explained, for example, by the dependence of beaming on apparent luminosity (objects that appear less luminous on average, like HBL, are less beamed), which could cause the (large!) K~correction to depend artificially and systematically on apparent luminosity (Giommi).

In contrast, quasars clearly evolve in the positive sense (Padovani, Jackson), and the highest luminosity BL Lacs might have similar evolution. In fact, a thorough analysis of radio sources spanning the full FR~I/II power range suggests there is a smooth dependence of evolution on radio power, with the low-power sources showing little or no evolution and the high-power sources showing large, positive 
evolution---in other words, luminosity-dependent evolution, similar to that seen in radio-quiet quasars (Jackson). An evolutionary connection between FSRQ and BL~Lacs has also been suggested (Stocke, D'Elia). This reinforces the previously mentioned ``grand unification'' of radio sources.

One can in principle extend the unification idea to even lower radio power, into the so-called ``radio-quiet'' regime. New, deeper radio surveys like the FIRST are finding a unimodal distribution of the radio-to-optical flux ratio---i.e.\ there do not appear to be two populations of radio sources, one radio-quiet and one 
radio-loud ---in contrast to what has been inferred from large optical samples 
(e.g.\ PG and LBQS). A close look at conventionally radio-quiet AGN indeed shows they have some blazar-like properties: flat-spectrum radio cores, large variability, high brightness temperatures, and in a handful of cases, superluminal expansion velocities (Falcke). {\it Thus AGN might all have similar nuclear engines, producing relativistic radio-emitting jets that in most cases fail to form large, bright radio
sources.} Given the demonstrated influence of (multiple) flux limits on BL~Lac samples, it is clearly past time to consider similar effects in quasar 
samples---perhaps the radio-quiet/radio-loud ``division'' is artificial too.

\section{Jet Physics}

The observed multiwavelength variability of blazars is complex and thus interpretation in terms of underlying jet physics is not so straightforward. For the high energy synchrotron radiation, both soft and hard lags are seen, in the same object, within the same sequence of short flares (Wagner). Soft lags can be explained by synchrotron cooling providing electron acceleration is rapid, while possibly hard lags can be explained by slow acceleration. In either case, the observed symmetry of flares implies the light crossing time is the dominant time scale, and the absence of prominent plateaus suggests that the electron injection time must be comparable to
it. Interestingly, the handful of blazars well studied with long ASCA observations show a break in the structure function at $\sim1$~day, similar to the radio and optical light curves of intra-day-variable AGN (Tanihata, Wagner). This rapid variability offers the strongest evidence for a characteristic time scale in any AGN. There appears to be a connection between X-ray and optical variability and 
$\gamma$-ray and radio variability, at least in some sources (Nesci, Jorstad).

Jets clearly carry considerable kinetic power, sufficient to power extended radio lobes in FR~II sources. How this energy is converted to radiation is not completely clear but it must be a relatively low-efficiency process (so as not to tap all the kinetic energy), and it may involve internal shocks, in the manner expected to hold in gamma-ray bursts (Ghisellini, Sikora, Cross). Far less clear is how the energy is extracted from the black hole and transported to the radiative zone. Intrinsic differences might exist in the inner jets of radio sources (Lister).

Most of the jet kinetic energy is probably carried by protons. If carried instead by electrons, there would have to be a large number of cold electrons (with minimum energy $\gamma_{\rm min}\sim1$). The bulk relativistic outflow in the jet ($\Gamma_{\rm bulk}\sim10$) would allow these electrons to Comptonize optical/UV photons to soft X-ray energies (Sikora). The absence of such a ``bulk Compton bump'' has been used to rule out such a low $\gamma_{\rm min}$; however, recent interpretations of soft X-ray spectra suggest such features may exist (Celotti). On the other hand, electron-positron jets are easier to slow down, and would probably be more consistent with the relatively slow jets observed, for example, in TeV BL Lacs (Marscher, Piner).  Whether or not protons dominate the kinetic energy budget, pairs certainly have to be present, if only because they are formed easily in high-energy-density
environments, through proton-proton collisions, proton-induced cascades, and 
$\gamma$$\gamma$ collisions (Sikora).

Chandra observations of radio jets will undoubtedly provide critical new information. Already there are some suprises. X-ray jets extend to very large scales, many tens of kiloparsecs in projected distance from the nucleus, up to half a Megaparsec 
(de-projected) in the most extreme example, PKS~0637$-$752. This emission is difficult to explain via synchrotron or synchrotron self-Compton models---the energy requirements are too large and/or the spectral shape implies multiple populations of electrons. A viable alternative is that relativistic electrons in the jet 
inverse-Compton scatter photons in the ambient cosmic microwave background radiation field, provided the jet still has relativistic bulk outflow on such large spatial scales (Tavecchio, Sambruna, Ghisellini). Further Chandra (and HST) observations will
test this important hypothesis. 

\section{Blazar Demographics}

The original goal of the conference was to address the demographics of blazars. A seemingly prosaic question---whether there are more ``red'' or ``blue'' 
blazars---actually has profound physical meaning since red blazars might be more powerful than blue, on average. That is, the demographics issue translates to the fundamental question: {\it What kind of jets does nature make, powerful or weak?}  This clearly bears on the radio-quiet issue as well: if radio-quiet quasars are much more numerous, and if (weak) jets are common in the nuclei of radio-quiet AGN, then weak jets have to be more common.

An answer to the demographics question proved elusive, largely due to the overwhelming influence of selection effects but also to the fact that most new
samples are still not completely identified. Before the information content of
those samples can be interpreted, extensive simulations will also be needed. Preliminary results presented at the conference make clear that quite broad
blazar populations---quite a range of jet physical states---are allowed by the current data. Physical limits (e.g.\ to peak synchrotron frequencies) are still not well constrained by any single sample. With joint constraints from various surveys sampling different regions of flux-flux space and improved simulations, however, it should be possible to definitively answer the demographics question. Thus interesting work still lies ahead.

\acknowledgments 

We acknowledge the Space Telescope Science Institute, via the AGN Journal Club, the Visitor Program, and the Director Discretionary Research Fund, for financial support. We thank Quindairian Gryce for expert organization of the meeting and Sharon Toolan for critical assistance with the proceedings. Finally, we enthusiastically thank our blazar colleagues for their valuable participation in this meeting. 


\end{document}